\documentstyle[aaspp4,12pt,amsmath]{article}
\def \sun {$_{\scriptscriptstyle\odot}$}

\lefthead{FRYER, HUNGERFORD, \& WARREN}

\righthead{$\gamma$-Ray Lines From Asymmetric Supernovae}

\received{2002 September 11}
\begin{document} 




\title{$\gamma$-Ray Lines From Asymmetric Supernovae}
\author{Aimee L. Hungerford$^{\textrm{1},\textrm{2}}$, Chris L. Fryer$^{\textrm{1}}$, Michael S. Warren$^{\textrm{1}}$} 
\affil{$^{\textrm{1}}$Los Alamos National Laboratories, \\ Los Alamos, NM 87544}
\affil{$^{\textrm{2}}$University of Arizona - Steward Observatory, \\ Tucson, AZ 85721}
\authoremail{aimee@lanl.gov}
\authoremail{fryer@lanl.gov}
\authoremail{mswarren@lanl.gov}

\begin{abstract}
We present 3-dimensional SPH simulations of supernova explosions from
100 seconds to 1 year after core-bounce.  By extending our modelling
efforts to a 3-dimensional hydrodynamics treatment, we are able to
investigate the effects of explosion asymmetries on mixing and
$\gamma$-ray line emergence in supernovae.  A series of initial
explosion conditions are implemented, including jet-like and
equatorial asymmetries of varying degree.  For comparison, symmetric
explosion models are also calculated.  A series of time slices from
the explosion evolution are further analyzed using a 3-dimensional
Monte Carlo $\gamma$-ray transport code.  The emergent hard X- and
$\gamma$-ray spectra are calculated as a function of both viewing
angle and time, including trends in the $\gamma$-ray line profiles.
We find significant differences in the velocity distribution of
radioactive nickel between the symmetric and asymmetric explosion
models.  The effects of this spatial distribution change are reflected
in the overall high energy spectrum, as well as in the individual
$\gamma$-ray line profiles.
\end{abstract}
\keywords{black hole physics --- supernovae: general ---stars: neutron --- gamma rays: theory}


\section{Introduction}

Among the many surprises that supernova (SN) 1987A brought astronomers
was the early emission of X-rays and $\gamma$-rays (X-rays:
e.g. Dotani et al.  1987; Sunyaev et al. 1987; $\gamma$-rays:
e.g. Cook et al. 1988; Mahoney et al. 1988; Matz et al. 1988).  This
high energy emission, arising from the decay of $^{56}$Co, appeared
nearly 6 months earlier than was predicted by theoretical models (e.g.
Pinto \& Woosley 1988a, Arnett et al. 1989) and led theorists to
conclude that the $^{56}$Ni, produced deep in the core of this
exploding star, had mixed into the outer layers of the supernova
ejecta (e.g.  Pinto \& Woosley 1988b, Arnett et al. 1989 and
references therein).  Additional evidence all seems to support the
mixing of SN~1987A's ejecta: models of the supernova light curves
(Arnett 1988; Shigeyama, Nomoto, \& Hashimoto 1988; Woosley 1988;
Shigeyama \& Nomoto 1990) and explanations of spectral line widths
(Haas et al. 1990; Spyromilio, Meikle, \& Allen 1990; Tueller et
al. 1990).  SN~1987A is not peculiar in this mixing.  Many supernovae
show evidence of mixing in their spectra (e.g.  Spyromilio 1994,
Fassia et al. 1998) and the light curves and spectra of Type Ib SN
seem to be best fit by mixed models (Shigeyama et al. 1990; Woosley \&
Eastman 1997).  It appears that mixing is a generic process in
core-collapse supernovae.

These results have stimulated a series of multi-dimensional
hydrodynamical simulations trying to produce the observed mixing
(Arnett, Fryxell, \& M\"uller 1989; Hachisu et al. 1990, Fryxell,
Arnett \& M\"uller 1991; M\"uller, Fryxell \& Arnett 1991; Herant \&
Benz 1991; Herant \& Benz 1992; Herant \& Woosley 1994; Kifonidis et
al. 2000; Kifonidis et al. 2003).  Although these simulations seem to
be able to explain the mixing in Type Ib supernovae (Kifonidis et al.
2000), none of these results are able to explain the extensive mixing
observed in SN~1987A.  Possible solutions to this shortcoming have
been proposed: (1) perhaps the decay of $^{56}$Ni injects enough
energy to force additional mixing (Herant \& Benz 1992) or (2)
convection in the pre-collapse core provides enough seeds to enhance
mixing (Herant \& Benz 1992).  A third possibility is that the
supernova explosion itself is asymmetric (Nagataki et al. 1997;
Nagataki et al.  1998; Nagataki 2000 and references therein).
Nagataki et al. (1998) found that not only could slight asymmetries in
the supernova explosion produce the required mixing to explain 1987A,
but they could also explain anomalies in the nucleosynthetic yields
produced by several supernovae.

The evidence for such global asymmetries in supernova explosions has
been growing steadily.  One piece of evidence arises from attempts to
understand the high space velocities of neutron stars.  The high
observed velocities of pulsars, along with evidence of neutron
star/remnant associations, and the formation scenarios of neutron star
binaries all suggest that neutron stars are given strong kicks at
birth.  These kicks are most easily explained by some asymmetry in the
supernova explosion (see Fryer, Burrows, \& Benz 1996 for a review).
In addition, the most straightforward explanation of the large
polarization seen in core-collapse supernovae (see Wang et al. 2001;
Leonard \& Filippenko 2001 and references therein) is that the
explosion driving these supernovae is inherently asymmetric (H\"oflich
1991).

The asymmetries are believed to have their origin in the explosion
mechanism itself.  For instance, even if the collapsing star is
initially spherically symmetric, some asymmetry can be produced due to
convection taking place in, and above, the proto-neutron star (Herant
et al. 1994; Burrows, Hayes, \& Fryxell 1995; Janka \& M\"uller 1996).
To date, these asymmetries are not extreme enough in the theoretical
models to explain the mixing.  Large asymmetries may occur if the
collapsing star is asymmetric due to nuclear burning (Burrows \& Hayes
1996; Lai 2000), however, sufficiently large departures from spherical
symmetry have only been produced by assuming extremely asymmetric
collapsing cores (Burrows \& Hayes 1996).  Alternatively, rotation can
produce strong asymmetries in the supernova explosion (M\"onchmeyer \&
M\"uller 1989; Janka \& M\"onchmeyer 1989; Fryer \& Heger 2000, 
Khokhlov et al. 1999).  The
nature of these asymmetries depends upon the angular momentum profile
of the collapsing star and, although most calculations predict jet-like
explosions along the rotation axis, some calculations imply that an
equatorial explosion could occur (M\"onchmeyer \& M\"uller 1989).

In this paper, we follow supernova explosions in 3-dimensions from
100~seconds to 1 year after core-bounce.  We model a series of initial
explosion conditions with both jet-like ``axial'' and equatorial
asymmetries of varying degree.  In \S 2, we describe these simulations
and their results with comparisons to past work.  In \S 3, we 
discuss the $\gamma$-ray emission from these explosions and present
calculations of $\gamma$-ray spectra as a function of time and viewing
angle.  We conclude with a disscussion of the observational prospects 
of these results, highlighting upcoming gamma-ray missions.

\section{Explosion Simulations}

For our hydrodynamic simulations, we have used the 15\,M\sun
progenitor (s15s7b) by Weaver \& Woosley (1993).  This star has been
evolved with a piston-driven explosion to 100\,s after bounce,
producing 0.24\,M\sun of $^{56}$Ni.  The total energy of this model is
roughly $1.5 \times 10^{51}$\,erg with roughly $1.0 \times
10^{51}$\,erg in kinetic energy.  As this explosion moves through the
star, the shock hits composition boundaries where strong entropy
gradients exist.  When the shock hits these boundaries,
Rayleigh-Taylor instabilities develop, which can grow and cause the
star to mix (Chevalier \& Klein 1978; Weaver \& Woosley 1980).  Our
simulations model this mixing and concentrate on the effects that
asymmetries have on it.

\subsection{Numerical Schemes}

We model a series of explosions from 100\,s to 1\,year after the
launch of the supernova shock (Table 1) using a 3-dimensional smooth
particle hydrodynamics code (see Warren et al. 2002 for details) based
on the parallel oct-tree algorithm developed by Warren \& Salmon
(1993).  This lagrangian code tracks the composition exactly.  For 2
of our simulations, we included the energy injection from $^{56}$Ni
and $^{56}$Co decay (and hence also trace the abundances of $^{56}$Co
and $^{56}$Fe).  For these 2 models, we assume that all of the decay
energy is injected into the gas.  At late times, this will
overestimate the total energy deposited, as a fraction of this energy
will escape, but since we would like an upper limit on the effects of
$^{56}$Ni decay, and because the injection of energy is less important
to the explosion dynamics at late times, this assumption is adequate.
For decay energy, we assume that the total energy from decay is $9.3
\times 10^{16}$\,erg\,g$^{-1}$ with 33\% being released from $^{56}$Ni
decay with a 6.1\,d half-life and the other 67\% being released from
$^{56}$Co decay with a 77\,d half-life (Colgate, Petschek, \& Kriese
1980).  This is similar to what Herant \& Benz (1992) and Herant \&
Woosley (1994) used for their models.  For an equation of state, we
use the same ``low-density'' equation of state developed by Blinnikov,
Dunina-Barkovskaya, \& Nadyozhin (1996) that is used in our progenitor
model.

We mapped the Weaver \& Woosley (1993) model with its 100\,s long,
spherically symmetric explosion into our 3D SPH code with 2.2 million
variably-massed particles.  We model the entire sphere, so there are
no axis boundaries in this simulation and the neutron star mass at the
center is added through an external gravitational force.  Asymmetries
are added artificially to the velocities only and, for most of the
explosions, are added in such a way to insure that the total kinetic
energy of the explosion is conserved.  The two asymmetries we model
are jet explosions:
\begin{equation}
v_{\rm radial}=(\alpha+\beta \times |z|/r) v_{\rm radial}^{\rm sym}
\end{equation}
and equatorial explosions:
\begin{equation}
v_{\rm radial}=(\alpha-\beta \times |x|/r)v_{\rm radial}^{\rm sym}
\end{equation}
where $v_{\rm radial}^{\rm sym}$ is the velocity from the one
dimensional calculation, $v_{\rm radial}$ is the radial velocity for
the asymmetric setup, and $x$, $z$, $r$ are the x-position,
z-position, and radius of the particle.  The values of $\alpha$ and
$\beta$ for each model are given in Table 1, along with the initial
kinetic energy of each model.  The magnitude of the asymmetries are
guided by the results of Fryer \& Heger (2000), who found that their
rotating core-collapse simulations in 2-dimensions had velocities that
were a factor of 2 higher in the pole than in the equator 1.5\,s after
bounce.  We do not use the extreme asymmetric explosions of Khokhlov
et al. (1999) which concentrate the explosion energy into a narrow
jet.  As we shall show (and as Nagataki 2000 argued), such extreme
asymmetries are not required to explain outward mixing of nickel in
supernovae.  Although at the end of their simulation, the asymmetry in
polar vs.  equatorial velocities is growing, the trend as the shock
moves outward will be to spherize the shock.  So at this point, it is
not clear whether the shock at 100\,s will be more or less asymmetric
than what Fryer \& Heger (2000) found at the end of their simulation.

After mapping these models into our 3-dimensional SPH code, we then 
run the explosion out to 1\,year.  When the shock reaches the edge 
of the star, we assume it is moving through a vacuum.  In reality, 
such a star will have a stellar wind atmosphere surrounding it, 
but the density of a 15\,M\sun wind is so low, that for the purposes 
of our simulation, zero density material is appropriate.  In 
addition, photon transport is not modeled in our calculations.  
However, prior to shock breakout, the photons are essentially trapped 
in the shock.  By the time the shock breaks out, most of the internal 
energy has already been converted to kinetic energy, so although 
including photon transport will change the ionization state of the 
exploding stellar material, it does not affect the kinematics 
significantly.

The smooth particle hydrodynamic technique automatically adds a level
of perturbation in the code.  With our initial conditions, these
perturbations are random with a maximum $1-\sigma$ deviation in a
shell of 5-7\%.  Such large deviations only occur at composition
boundaries, specifically the silicon and oxygen shells, and match well
the deviations arising from explosive oxygen and silicon flashes prior
to collapse (e.g. Bazan \& Arnett 1998).

\subsection{Explosion and Nickel Distribution}

Although we use the same 15\,M\sun progenitor that was used by Herant
\& Woosley (1994), it has a higher explosion energy and we map this
model onto our 3-dimensional grid 200\,s earlier than Herant \&
Woosley (1994) mapped their spherically-symmetric explosion simulation
onto a 2-dimensional grid.  Therefore, although their study has the
closest similarities with our work, it is difficult to make direct
comparisons to their simulations.  Nevertheless, it is interesting to
compare the velocity distribution of each chemical element from our
3-dimensional simulations with the 2-dimensional simulations of Herant
\& Woosley (1994) at 90 days (compare the lower right panel in Fig. 1
of Herant \& Woosley 1994 with Fig. 1 in this paper).  In Fig.1, the
material labelled ``hydrogen'' includes all material in the hydrogen
envelope (as did Herant \& Woosley 1994).  Similarly, by ``nickel'' we
refer to both the distribution of nickel as well as its decay products
(most notably $^{56}Co$).  Although the distribution of elements is
similar in both the 2 and 3 dimensional simulations, the stronger
3-dimensional explosion causes all of the ejecta to be moving slightly
faster than that of the 2-dimensional simulation and it is difficult
to compare mixing instabilities.

Comparing the convective instabilities themselves is also difficult.
In 3-dimensions, the ``mushroom''-like structures formed by
Rayleigh-Taylor instabilities are not so well defined, and don't lie
along any one plane.  However, 4.3 hours into the explosion, it is
clear that instabilities have developed (Fig. 2) and these
instabilities ultimately mix nickel knots far out into the star
(Fig. 3).  At the start of the explosion, $^{56}$Ni is found only in
the inner 1.6\,M\sun of the star (1.3\,M\sun becomes the neutron star,
so the nickel is limited to the inner 0.3\,M\sun of ejecta).  By the
end of the simulation, $^{56}$Ni has mixed out nearly to 5\,M\sun,
beyond the $\sim$4.5\,M\sun boundary that marked the edge of the
helium layer (Fig. 4).  Unfortunately, this mixing is, if anything,
less than the amount of mixing found in the 2-dimensional simulations
of Herant \& Woosley (1994).  The fact that the mixing is less in
3-dimensions vs. 2-dimensions could be due to the lower effective
resolution (we only have 2.2 million particles in 3-dimensions
vs. 25,000 particles in the 2-dimensional simulations)\footnote{Note
that the 2-dimensional simulations of Herant \& Woosley (1994) have
poor mass resolution and low resolution could be a problem in the
2-dimensional simulations as well as the 3-dimensional simulations}.
However, bear in mind that the turbulent inverse cascade behaves 
differently in 2 and 3 dimensions (2-dimensional inverse cascades 
drive energy to large scales whereas 3-dimensional simulations suggest 
the energy is driven to small scales and dissipated).  It is likely 
that these differences cause the 2-dimensional simulations to produce 
more extended instabilities.  In any event, it appears that neither our
spherical 3-dimensional simulations nor the 2-dimensional simulations
seem to give enough mixing to explain the observations of supernovae
like 1987A.

Nagataki et al. (1998) and Nagataki (2000) found that they required
mild asymmetries ($v_{\rm Pole}/v_{\rm Equator}=2$) to explain
SN~1987A.  Since we model a 15\,M\sun star, not a SN~1987A progenitor,
it is difficult to both compare with this past work as well as
constrain our results with observations of SN~1987A.  But we can
discuss the basic trends caused by asymmetries.  Fig. 5 shows model
Jet2 1 year after explosion.  Note that although the density
distribution has spherized as the shock propogates through the shallow
density gradients of the red supergiant envelope (Chevalier \& Soker
1989), the distribution of $^{56}$Co (the decay product of $^{56}$Ni)
retains a large asymmetry.  We discuss the effects of these
asymmetries on the $\gamma$ ray emission in \S 3.

Like Nagataki et al. (1998) and Nagataki (2000), we find that the
asymmetries broaden the velocity profile of $^{56}$Ni (Fig. 6).
However, in our simulations, mild asymmetries ($v_{\rm Pole}/v_{\rm
Equator}=2$) led to only a small increase in th maximum nickel
velocity from 2500\,km\,s$^{-1}$ to 2900\,km\,s$^{-1}$.  For Nagataki
et al. (1998), such mild asymmetries increase the maximum nickel
velocity from 2200\,km\,s$^{-1}$ to 3200\,km\,s$^{-1}$!  This
difference could be progenitor dependent, an effect of 3-dimensional
vs.  2-dimensional convection, or the lack of resolution in our
3-dimensional models.  Extracting the true cause of this difference
awaits future calculations with similar initial conditions.

However, increasing the amount of asymmetry by another factor of 2
(Models Jet4, Eq4) causes some nickel to be ejected at very high
velocities.  The amount of mixing in these cases reaches extremes with
the nickel well into the hydrogen envelope.  Such mixing has decided
signatures in both the emergence of the $\gamma$-ray line flux and the
shape of the $\gamma$-ray lines (\S 3).  Note that the energy released
from the decay of nickel also helps to mix out the nickel (on par with
the effects of mild asymmetries).  Clearly, the energy released from
the decay of nickel cannot be neglected in any accurate mixing
calculation.

This mixing also has important repercussions for nucleosynthetic
yields and the mass-cut for the remnant mass.  Most black holes are
formed in stars which produce supernova explosions that are too weak to
throw off all of the stellar envelope and the subsequent fallback
produces a black hole (Fryer \& Kalogera 2001).  In our models (Jet2,
Sym+Decay, etc.), more than 10\% of the nickel produced is ejected
well beyond the helium core (Fig. 7).  If this trend holds for more
massive stars such as the progenitor of SN 1997D (Turatto et
al. 1998), then the entire helium core of such a star ($>$8\,M\sun)
could fall back and still enough nickel would escape to power the
observed light curve!  Bear in mind, however, that weaker explosions
may well produce less mixing, so adapting the results of our
simulations to supernovae like 1997D must be taken with some caution.

\clearpage
\begin{deluxetable}{lcccccc}
\tablewidth{38pc} 
\tablecaption{Explosion Simulations} 
\tablehead{
\colhead{Model\tablenotemark{a}} & 
\colhead{$V_{\rm pole}/V_{\rm equator}$} & 
\colhead{$\alpha,\beta$\tablenotemark{b}} & 
\colhead{K.E.$^{Initial}$} &
\colhead{T.E.$^{Initial}$} & \colhead{K.E.$^{final}$} & 
\colhead{Mixing\tablenotemark{c}} \\ \colhead{} & 
\colhead{} & \colhead{} & \colhead{($10^{51}$ergs)} & 
\colhead{($10^{51}$ergs)} & \colhead{($>$M\sun)}}

\startdata

Sym  & 1.0 & 1,0 & 1.0 & 0.51 & 1.3 & 4.6 \\
Sym D  & 1.0 & 1,0 & 1.0 & 0.51 & 1.3 & 5.5 \\
Jet2  & 2.0 & $\sqrt{3/7},\sqrt{3/7}$ & 1.0 & 0.51 & 1.3 & 5.6 \\
Jet2 D & 2.0 & $\sqrt{3/7},\sqrt{3/7}$ & 1.0 & 0.51 & 1.3 & 5.7 \\
Jet4  & 4.0 & $\sqrt{1/7},\sqrt{9/7}$ & 1.0 & 0.51 & 1.3 & 8.0 \\
Eq2  & 0.5 &  $4/3,2/3$ & 1.0 & 0.51 & 1.3 & 5.0 \\
Eq4  & 0.25 & $8/5,6/5$ & 1.2\tablenotemark{d} & 0.51 & 1.6 & 11.3 \\

\enddata

\tablenotetext{a}{The models include ones with symmetric initial
conditions (Sym), explosions along the polar axis (Jet), and
explosions along the equator (Eq).  Those models which include the
effects of Nickel and Cobalt decay are denoted with a ``D'' suffix.}
\tablenotetext{c}{For the polar explosions, the radial velocity
($v_{\rm radial}$) is given by: $v_{\rm radial}=(\alpha+\beta \times
|z|/r) v_{\rm radial}^{\rm sym}$ and for equatorial explosions:
$v_{\rm radial}=(\alpha-\beta \times |x|/r)v_{\rm radial}^{\rm sym}$
where $v_{\rm radial}^{\rm sym}$ is the velocity from the one
dimensional calculation, $x$, $z$, $r$ are the x-position, z-position,
and radius of the particle.}
\tablenotetext{c}{We limit the extent of mixing by the  
furthest position (in mass coordinates) beyond which more than 
3\% of the nickel produced near the core is mixed out.}  
\tablenotetext{d}{Note that the initial energy for the extreme 
asymmetric model was larger than the other models.  In part, 
this explains the extended mixing of this model.}

\end{deluxetable}
\clearpage

\section{High Energy Spectral Calculations}

For our spectral calculations, we have used data from the
3-dimensional explosion simulation discussed in the previous section.
We input ejecta material properties from five different snapshots in
time at 150, 200, 250, 300 and 365 days after explosion.  Spectral
calculations were carried out for both the Jet2 and Symmetric
explosion models.  Our analysis of these model spectra concentrates on
the differences in total luminosity and line profile shape with the
introduction of realistic explosion asymmetries.  Since the progenitor
star used as input to our simulations was a 15~M\sun red supergiant,
we are unable to directly compare our calculated spectra with the
observed high energy spectra of SN~1987A.  However, we discuss how our
models compare to various spectral trends observed from SN~1987A.

\subsection{Numerical Schemes}

We used a Monte Carlo technique, similar to that described in Ambwani
\& Sutherland 1988, for modelling $\gamma$-ray transport in
3-dimensions.  Input models of the supernova ejecta (element
abundances, density and velocities) were taken from the ``Jet2'' and
``Symmetric'' SPH explosion simulations and mapped onto a
140~$\times$140~$\times$140 cartesian grid.  Escaping photons were
tallied into 250 coarse energy bins, with finer binning at the decay
line energies to provide line profile information.  The emergent
photons were also tallied into 11 angular bins
($\Delta\theta$~=~10$^\circ$) along the polar axis (the models
investigated in this work are axisymmetric, alleviating the need to
tally in azimuthal angle as well.)

The decay of the radioactive species (predominantly $^{56}$Ni and its
decay product: $^{56}$Co) in the supernova ejecta gives rise to the
$\gamma$-ray line emission.  As in Ambwani \& Sutherland 1988, we
assign the energy of the emitted photon packets according to the decay
probabilities (Lederer \& Shirley 1978) for the various radioactive
species ($^{56}$Ni, $^{56}$Co, $^{57}$Co, $^{44}$Ti, $^{44}$Sc, and
$^{22}$Na).  We include a total of 56 decay lines from these species,
but for the explosion times considered, the packets fall predominantly
into $\sim$~15 decay lines.  Roughly $10^9$ Monte Carlo photon bundles
were generated for each input model in proportion to the mass of
radioactive material distributed throughout the ejecta.  The material
properties of the ejecta were not evolved with photon flight time.
However, we found that 99~\% of the photons contributing to the
observed model spectra have escape times of less than 2 days.  This is
sufficiently shorter than the timescales for change in the
hydrodynamic models that our assumption of a fixed material background
should be valid for the time slices considered here.  The luminosity
weight of each photon packet and the opacities seen by the packet were
calculated in the comoving frame of the fluid, but all photon
properties were boosted to the observer's frame before being tallied
into spectral observables.

In all models, photoelectric and pair production opacities were
calculated for the elements H, He, C, N, O, Ne, Mg, Si, S, Ar, Ca, Ti,
Cr, Fe, Co, and Ni which correspond to the elements used in the
nucleosynthesis calculations for the progenitor star from Weaver \&
Woosley (1993).  The cross section data for these elements were taken
from the LLNL Evaluated Nuclear Data Library (Plechaty, Cullen, and
Howerton 1981, revised 1987).  The angle- and energy-dependent Compton
scattering opacities were calculated assuming that all electrons,
bound and free, contribute to the total cross section.  Daughter
products from the absorption processes were not followed
(e.g. positron annihilation photons from pair production and K-shell
fluorescence photons from photoelectric absorption.)  In order to test
the validity of this last approximation, we compared the results of
our 3-dimensional Monte Carlo transport code (Maverick) with the
1-dimensional code FASTGAM (Pinto \& Woosley 1988a).  Good agreement
was found between the model spectra whether the daughter products of
absorption were included in FASTGAM or not.  This justified the
decision to disregard the fluorescence and annihilation photons in
Maverick.

Figure 8, shows comparison spectra with the 1-dimensional FASTGAM code
for model 10HMM (Pinto \& Woosley 1988b).  This is a logarithmic plot
of photon flux (photons/second/MeV/cm$^2$, assuming a distance of
60~kpc) across the energy range 0.3~keV - 4~MeV.  The input model
(10HMM) is an artificially mixed version of the Woosley 1988 10H model
progenitor for SN~1987A.  The first comparisons between FASTGAM and
Maverick (left panel of Figure 8) showed significant differences in
the location of the hard X-ray fall off.  This was due to an invalid
assumption in FASTGAM of constant electron fraction for all species
which contribute to photoelectric and pair production opacities.  This
assumption is not valid for hydrogen, and results in a lower
calculated opacity for the lower energy X-rays.  The right panel of
Figure 8 shows the hard X- and $\gamma$-ray spectra calculated from
the two different codes after the absorption opacity correction was
made.  There is good agreement across the spectrum to within the
uncertainties of the Monte Carlo calculation.

\subsection{Hard X-ray and $\gamma$-ray Spectrum}

Figure 9 is a logarithmic plot of photon flux in units of
photons/second/MeV/cm$^2$ across the energy range investigated with
these simulations (0.3~keV - 4~MeV).  We have placed this object at
the distance of the Large Magellanic Cloud (60~kpc) for easy
comparison with flux data from SN~1987A observations.  The 5 panels
are spectra from the different time slices; in each panel, we plot the
spectrum for the Symmetric model, along with polar and equatorial
views of the Jet2 model.  It can be seen immediately that the hard
X-rays emerge earlier from the ejecta with a global explosion
asymmetry (Jet2 model).  This holds regardless of viewing angle (pole
versus equator) towards the explosion.

The fact that the hard X-ray flux in the aspherical explosion model is
larger than the symmetric explosion, regardless of line of sight, can
be understood in principle from optical depth arguments.  In Figure
10, we show a contour plot of density (outer contour) and $^{56}$Co
number density (inner contour) for the Jet2 and Symmetric models at
t~=~150~days.  Decay of $^{56}$Co is the major source of $\gamma$-ray
photons, so the inner contour essentially traces the surface of the
emission region.  The horizontal and vertical lines in Figure 10
represent lines of sight from the ejecta surface to the emission
source and are labeled with the optical depth along that line of
sight.  The dominant opacity for the hard X- and $\gamma$-rays is
Compton scattering off electrons and, since the density contours
remain roughly spherical in both models, the optical depth from a
given point to the ejecta surface is roughly constant.  In the Jet2
model, the $^{56}$Ni was mixed out to larger radii in the polar
direction, so it makes perfect sense that we see enhanced emission
over the Symmetric model spectrum for that viewing angle (the optical
depth that the high energy photons must pass through is 10 in the
symmetric model versus $\sim 6-7$ along the polar line-of-sight in the
Jet2 model).

At a first glance, one might expect that, in the Jet2 explosion along
the equatorial line-of-sight, the total flux should also be low (the
optical depth from the nickel in the equator is also roughly 10).
However, this material does not dominate the high energy emission seen
along the equatorial line-of-sight.  The material ejected along the
poles has been mixed far enough out in the ejecta that the optical
depth these high energy photons must travel through, even along the equatorial
line-of-sight, is quite low ($\sim 7$).  It is this nickel which
dominates the hard X-ray emission at all viewing angles.  In fact, the
optical depth from the ``ends'' of the $^{56}$Co distribution does not
differ very much between the polar view and the equator view
($\tau$~=~6~\&~7 respectively), which explains why the overall hard
X-ray flux depends only mildly on viewing angle.

For the later time slices, this mismatch in escaping emission from the
``ends'' versus the equatorial plane ejecta is less pronounced, and
the equator view spectrum has comparable contributions from both
regions.

\clearpage
{\tiny

\begin{deluxetable}{lc@{\hspace{0.2cm}}c@{\hspace{0.2cm}}c@{\hspace{0.2cm}}c@{\hspace{0.2cm}}c@{\hspace{0.2cm}}c@{\hspace{0.2cm}}c@{\hspace{0.2cm}}c}
\tablewidth{44pc} 
\tablecaption{High Energy Luminosities\tablenotemark{a}} 
\tablehead{
\colhead{Model\tablenotemark{b}} & 
\colhead{Time} & 
\colhead{3-30~keV\tablenotemark{c}} & 
\colhead{30-100~keV\tablenotemark{c}} & 
\colhead{100-500~keV\tablenotemark{c}} & 
\colhead{500-1000~keV\tablenotemark{c}} & 
\colhead{1000-3800~keV\tablenotemark{c}} & 
\colhead{847~keV Line\tablenotemark{d}} & 
\colhead{1238~keV Line\tablenotemark{d}}\\
    & \colhead{(day)} &  &  & & & & & 
}

\startdata

Sym &  &  &  &  &  &  &  & \\
              & 150 & 3.715(0.138)$\times$10$^{35}$ 
                    & 2.008(0.162)$\times$10$^{36}$  
                    & 6.537(0.930)$\times$10$^{36}$ 
                    & 9.281(2.135)$\times$10$^{36}$ 
                    & 3.973(1.448)$\times$10$^{37}$ 
                    & 6.828(1.593)$\times$10$^{41}$ 
                    & 4.687(1.588)$\times$10$^{41}$ \\
              & 200 & 5.172(0.044)$\times$10$^{36}$ 
                    & 3.183(0.055)$\times$10$^{37}$ 
                    & 1.058(0.032)$\times$10$^{38}$ 
                    & 1.440(0.075)$\times$10$^{38}$ 
                    & 5.685(0.498)$\times$10$^{38}$ 
                    & 1.156(0.061)$\times$10$^{43}$ 
                    & 9.622(0.632)$\times$10$^{42}$  \\
              & 250 & 1.844(0.006)$\times$10$^{37}$
                    & 1.272(0.009)$\times$10$^{38}$ 
                    & 4.302(0.052)$\times$10$^{38}$ 
                    & 5.727(0.119)$\times$10$^{38}$ 
                    & 2.015(0.074)$\times$10$^{39}$ 
                    & 5.008(0.103)$\times$10$^{43}$
                    & 4.476(0.106)$\times$10$^{43}$\\
              & 300 & 3.714(0.007)$\times$10$^{37}$
                    & 2.996(0.011)$\times$10$^{38}$
                    & 1.043(0.007)$\times$10$^{39}$
                    & 1.367(0.015)$\times$10$^{39}$
                    & 4.257(0.085)$\times$10$^{39}$
                    & 1.362(0.014)$\times$10$^{44}$
                    & 1.280(0.014)$\times$10$^{44}$\\
              & 365 & 4.529(0.010)$\times$10$^{37}$
                    & 4.910(0.018)$\times$10$^{38}$
                    & 1.855(0.011)$\times$10$^{39}$
                    & 2.446(0.025)$\times$10$^{39}$
                    & 6.650(0.131)$\times$10$^{39}$
                    & 3.016(0.027)$\times$10$^{44}$
                    & 2.886(0.026)$\times$10$^{44}$\\

Jet2-E &  &  &  &  &  &  &  & \\
              & 150 & 1.503(0.585)$\times$10$^{36}$ 
                    & 9.891(4.221)$\times$10$^{36}$  
                    & 2.812(1.573)$\times$10$^{37}$ 
                    & 3.877(2.897)$\times$10$^{37}$ 
                    & 1.200(1.089)$\times$10$^{38}$ 
                    & 2.748(2.547)$\times$10$^{42}$ 
                    & 3.041(2.813)$\times$10$^{42}$ \\
              & 200 & 8.262(1.177)$\times$10$^{36}$ 
                    & 5.565(1.103)$\times$10$^{37}$ 
                    & 1.779(0.543)$\times$10$^{38}$ 
                    & 2.066(1.004)$\times$10$^{38}$ 
                    & 6.171(4.649)$\times$10$^{38}$ 
                    & 2.150(1.188)$\times$10$^{43}$ 
                    & 1.244(0.942)$\times$10$^{43}$  \\
              & 250 & 2.783(0.164)$\times$10$^{37}$
                    & 2.290(0.249)$\times$10$^{38}$ 
                    & 7.969(1.379)$\times$10$^{38}$ 
                    & 8.943(2.822)$\times$10$^{38}$ 
                    & 2.700(1.366)$\times$10$^{39}$ 
                    & 8.641(2.862)$\times$10$^{43}$
                    & 7.455(2.908)$\times$10$^{43}$\\
              & 300 & 4.260(0.279)$\times$10$^{37}$
                    & 3.996(0.460)$\times$10$^{38}$
                    & 1.475(0.262)$\times$10$^{39}$
                    & 1.774(0.553)$\times$10$^{39}$
                    & 5.049(2.500)$\times$10$^{39}$
                    & 1.937(0.612)$\times$10$^{44}$
                    & 1.712(0.572)$\times$10$^{44}$\\
              & 365 & 4.576(0.218)$\times$10$^{37}$
                    & 5.667(0.435)$\times$10$^{38}$
                    & 2.285(0.268)$\times$10$^{39}$
                    & 2.933(0.613)$\times$10$^{39}$
                    & 7.678(2.686)$\times$10$^{39}$
                    & 3.864(0.674)$\times$10$^{44}$
                    & 3.624(0.634)$\times$10$^{44}$\\

Jet2-P &  &  &  &  &  &  &  & \\
              & 150 & 2.136(0.710)$\times$10$^{36}$
                    & 1.447(0.516)$\times$10$^{37}$
                    & 4.529(2.289)$\times$10$^{37}$
                    & 7.679(4.313)$\times$10$^{37}$
                    & 3.592(2.866)$\times$10$^{38}$
                    & 5.486(3.883)$\times$10$^{42}$
                    & 3.041(2.820)$\times$10$^{42}$\\
              & 200 & 1.160(0.139)$\times$10$^{37}$
                    & 8.387(1.527)$\times$10$^{37}$
                    & 2.887(0.783)$\times$10$^{38}$
                    & 4.747(2.002)$\times$10$^{38}$
                    & 1.904(1.098)$\times$10$^{39}$
                    & 4.442(1.973)$\times$10$^{43}$
                    & 4.628(2.118)$\times$10$^{43}$\\
              & 250 & 3.461(0.181)$\times$10$^{37}$
                    & 3.191(0.301)$\times$10$^{38}$
                    & 1.186(0.181)$\times$10$^{39}$
                    & 1.824(0.441)$\times$10$^{39}$
                    & 5.588(2.098)$\times$10$^{39}$
                    & 2.244(0.439)$\times$10$^{44}$
                    & 2.334(0.448)$\times$10$^{44}$\\
              & 300 & 4.556(0.286)$\times$10$^{37}$
                    & 5.015(0.520)$\times$10$^{38}$
                    & 1.995(0.319)$\times$10$^{39}$
                    & 2.987(0.782)$\times$10$^{39}$
                    & 8.274(3.336)$\times$10$^{39}$
                    & 4.357(0.884)$\times$10$^{44}$
                    & 4.210(0.821)$\times$10$^{44}$\\
              & 365 & 4.151(0.208)$\times$10$^{37}$
                    & 6.217(0.464)$\times$10$^{38}$
                    & 2.746(0.300)$\times$10$^{39}$
                    & 4.048(0.722)$\times$10$^{39}$
                    & 1.010(0.312)$\times$10$^{40}$
                    & 7.200(0.899)$\times$10$^{44}$
                    & 6.383(0.832)$\times$10$^{44}$\\

\enddata

\tablenotetext{a}{Numbers in parentheses represent $1-\sigma$ 
                  Monte Carlo uncertainties.}
\tablenotetext{b}{The input models include the Symmetric 
                  explosion model (Sym), the Jet2 explosion 
                  model for the polar (Jet2-P) and equator 
                  (Jet2-E) viewing angles}
\tablenotetext{c}{Units are erg~s$^{-1}$}
\tablenotetext{d}{Units are $\gamma$~s$^{-1}$}

\end{deluxetable}
}
\clearpage

\subsection{$\gamma$-ray Line Profiles}

Although the overall hard X-ray emission shows little variation
between pole and equator views, a detailed look at the $\gamma$-ray
line profile shapes and strengths, for the 1.238 and 0.847 MeV
$^{56}$Co lines, does reveal trends with viewing angle.  Figures 11
and 12 show line profiles of these two $^{56}$Co lines for both the
Symmetric and Jet2 explosion models.  The broadening of the line is
caused by Doppler velocity shifts resulting from the spatial 
distribution of radioactive nickel in the homologously expanding
ejecta.  The 4 panels shown are for days 200, 250, 300, and 365 after
explosion.  The three lines in the Jet2 spectra represent different
viewing angles through the ejecta (along the pole, the equator and an
intermediate angle $\sim 45^\circ$.)  For the Symmetric spectra, we
have plotted these same viewing angles.

\subsubsection{Global Asymmetry}

Both explosion scenarios (Symmetric and Jet2) show blue-shifted line
profiles and this effect is most enhanced in the Jet2 model.  These
differences can be best understood by examining the physical effects
which dictate the formation of the line profile edges.  The blue edge
to the lines is set by the maximum observed line-of-sight velocity of
the $^{56}$Co in the ejecta.  Because the nickel was mixed further out
(and achieves higher velocities) for the asymmetric explosions, the
$\gamma$-ray line profiles viewed along the line-of-sight of the jet are
much more blueshifted for the Jet2 model than the Symmetric model

Note, however, that the line profiles in the asymmetric explosion
model (Jet2) depend upon viewing angle.  $90^{\circ}$ off the jet
axis, the nickel producing the line profiles arises from both
fast-moving jet ejecta and the ejecta along the equator.  Because the
fast-moving ejecta is moving perpendicular to the line of sight, the
$\gamma$-rays from this material are not blueshifted significantly,
and the slow moving equatorial material also produces only a modest
blueshift.  Hence, the line profiles viewed along the equatorial
line-of-sight are blueshifted much less than those viewed along the
polar line-of-sight.

The red edge of the lines is determined by the escaping emission from
$^{56}$Co with the smallest line-of-sight velocity in the ejecta.  In
a Symmetric model, this should be an indication of how deep into the
ejecta we can see along a given viewing angle.  However, there is a
more pronounced effect at play in the asymmetric explosion models.  As
we mentioned in the previous section, much of the $\gamma$-ray
emission for the equatorial view arises from the ``ends'' of the $^{56}$Co
distribution.  This material has a very low line-of-sight velocity for
an equatorial observer, since it is being ejected predominantly in the
polar direction.  This allows for a significantly lower velocity red
edge of the equator view lines, even though the optical depth profiles
do not vary much between polar and equator viewing angles.

Also note that the line centroids become less blue-shifted with time.
As the supernova expands, the opacity in the ejecta drops and emission
from material located deeper into the ejecta (and thus at smaller
radial velocities) becomes visible.  This results in an overall
broadening of the line, as well as a redward shift of the line
centroid.

\subsubsection{Clumping Asymmetry}

This basic understanding of the line profile edges addresses
only global features of the emitted decay lines.  To understand the range
of possibilities that the line shape may take between those edges, we must
concentrate on the spatial distribution of the radioactive elements in the
supernova ejecta.

Because the expansion is basically homologous after 100 days, the
line-of-sight velocity of a fluid element in the ejecta is
proportional to its distance above the mid-plane of the explosion.
Each spectral energy bin in the line profile can be mapped to a unique
line-of-sight velocity in the ejecta, which can in turn be mapped to a
specific height above the mid-plane.  So, given the distribution of
cobalt as a function of x, y and z-coordinates (~${\mathcal
N}_{\mathrm{Co}}(x,y,z)$\ , where z is along the line-of-sight to the
observer), the line profile shape should be proportional to

\begin{equation}
{\mathbf N}_{\mathrm{Co}}(z) = \sum_{x,y} {\mathcal N}_{\mathrm{Co}}(x,y,z).
\end{equation}

Optical depth effects, as discussed above, will alter the line shape
from this basic picture, but it is important to realize that the
clumping, which occurs as the inner material is mixed outward, can
cause sizeable irregularities in the profiles.  Figure 13 shows a plot
of ${\mathbf N}_{\mathrm{Co}}$ versus line-of-sight velocity for the
polar view of model Jet2 at 250 days with higher spatial resolution
than the previous run.  The emergent $\gamma$-ray line flux is plotted
with a dotted line and can be seen to peak at velocities of around
1800~km/s.  This corresponds to the location in the ejecta at which
the exponential attenuation from the optical depth surpasses the flux
enhancement from increasing cobalt mass: $\tau_{MeV}\sim 4$.  As the
ejecta expand and optical depth drops, the $\gamma$-ray line profiles
will take on the irregular shape seen in the cobalt distribution.

\section{Conclusions}

In our simulations, we found that asymmetric explosions lead to
extensive mixing of the supernova ejecta, placing the products
of explosive burning well into the helium layer of the star.  Even
mild 2:1 asymmetries can mix 10\% of the nickel out to the inner edge
of the hydrogen layer.  If such mixing occurred in weak explosions, as
well as the strong explosions presented in this paper, these mildly
asymmetric supernova explosions could explain the extensive mixing
required in population III stars (Umeda \& Nomoto 2002) and black hole
binary systems such as Nova Scorpii (Podsiadlowski et al. 2002).

From the Monte Carlo transport simulations, we find that the high
energy fluxes at early times ($\sim$150 days) for the asymmetric
explosion model is roughly 4-5 times larger than for the corresponding
symmetric explosion model.  This suggests that emission from an
explosion with mixing {\it and} global asymmetry would be observed
roughly 50-100 days earlier than a symmetric explosion model with
mixing alone.  In addition, the line profiles for the asymmetric
explosion vary with viewing angle and are brightest for the polar
view.  Along this viewing angle, the integrated line flux is enhanced
by a factor of 2-3 over the symmetric explosion, and the lines peak at
roughly 3-5 times brighter than the symmetric explosion line profiles.
The integrated flux enhancement for the equator view of the Jet2 model
is of order 30~\% over the symmetric model.  The line centroids
observed for the asymmetric explosion are shifted relative to those
from the symmetric model (more to the blue for the polar view and
slightly more to the red for the equator view.)  At later epochs in
the SN explosion, the line profiles should reflect the irregular
shapes seen in the underlying cobalt distribution, which result from
the clumpy nature of the outward mixing of nickel in the ejecta.
Table~2. summarizes the calculated luminosities of the
different models for various continuum energy bands and the 1.238 and
0.847~MeV $^{56}$Co lines.

With the recent launch of the International Gamma-Ray Astrophysics
Laboratory (INTEGRAL), it is particularly interesting to analyze our
results in terms of the instrumental resolution and sensitivities this
satellite can provide across these energy bands.  It is important to
keep in mind that the nickel mass synthesized in this 15\,M\sun \  model
is roughly 3 times larger than the typically observed value
($\sim$0.08\,M\sun) for core-collapse SN explosions.  Since the high
energy emission should scale linearly with nickel mass, the fluxes
observed for this model will be roughly 3 times brighter than we would
expect for a more typical core-collapse event.  In fact, the high
energy continuum for our Symmetric model explosion is of order 3 times
brighter than was observed towards SN~1987A.  Since 1987A's nickel
yield was fairly typical, this comparison seems to validate the linear 
scaling of flux with nickel mass.  

Our models suggest that $\gamma$-ray line observations will be most 
useful for diagnosing the departure of a core-collapse explosion from 
spherical symmetry.
At energies around 1~MeV, INTEGRAL will have a spectral resolution of
2~keV and a narrow line sensitivity ($3-\sigma$ in 10$^{6}$~seconds) of 
$\sim$5$\times$10$^{-6}$~phot~s$^{-1}$~cm$^{-2}$ (Hermsen \& Winkler 2002).
Our model lines are about 5 times broader than this resolution element, so
the sensitivity for detecting them is worse by roughly $\sqrt{5}$.  Using these
specifications, INTEGRAL would be able to detect the $^{56}$Co lines
from the polar view of the Jet2 model at a distance of $\sim$650~kpc.

\acknowledgements We would like to thank P. Pinto, T. Evans,
T. Urbatsch, A. Heger, K. Nomoto, S. Woosley, J. Cannizzo and
N. Gehrels for their advice and encouragement on this paper.  The
large hydrodynamical simulations were run on the Avalon cluster at
LANL.  This work was performed under the auspices of the
U.S. Department of Energy by Los Alamos National Laboratory under
contract W-7405-ENG-36.  Funding was also provided by a Feynman
Fellowship at LANL, DOE SciDAC grant number DE-FC02-01ER41176 and
Beomax, Inc.  These simulations were run on the Avalon and
Space-Simulator clusters at LANL.

\clearpage
\begin{figure}
\plotfiddle{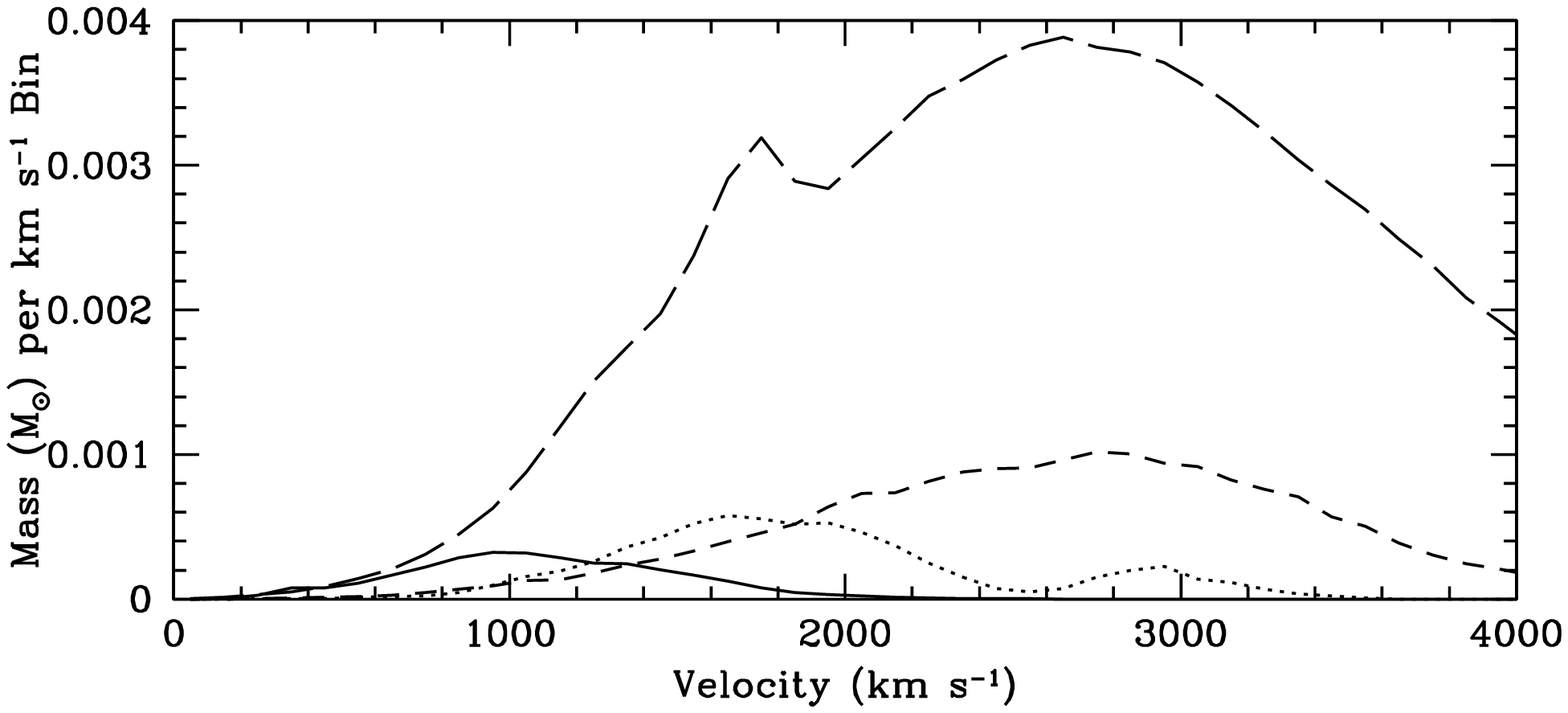}{7in}{0}{80}{80}{-250}{-10}
\label{fig:element_distrib}
\caption{Velocity distribution of nickel (solid), Oxygen (dotted),
Helium (dashed), and Hydrogen (long dashed) in our 3-dimensional
simulations.  Comparing these distributions to the 2-dimensional
simulations in Fig. 1 of Herant \& Woosley (1994), we note that the
distribution of elements is similar in both the 2- and 3-dimensional
simulations.  However, the stronger 3-dimensional explosion causes all
of the ejecta to be moving slightly faster than that of the
2-dimensional simulation and it is difficult to compare mixing
instabilities.}
\end{figure}
\clearpage

\begin{figure}
\plotfiddle{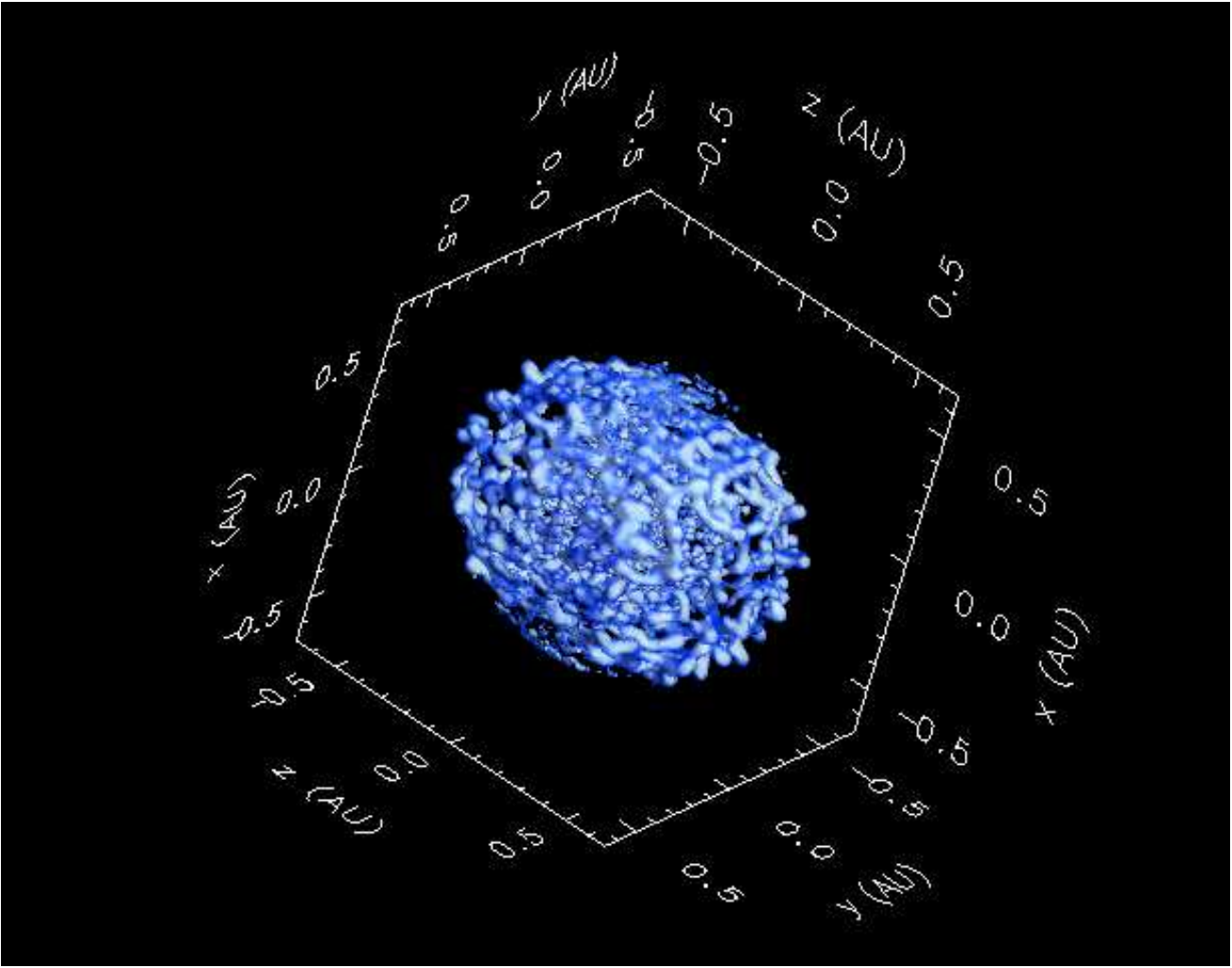}{7in}{0}{80}{80}{-250}{-10}
\label{fig:co_tau.ps_a}
\caption{Density contour (7$\times$10$^{-5}$~g\,cm$^{-3}$) plot of the early
  stages of the convection, 4.3 hours after the launch of the
  explosion.  Notice that tendrils mixing out the material have
  already developed.  It is this mixing that places nickel far beyond
  its initial distribution.}
\end{figure}
\clearpage

\begin{figure}
\plotfiddle{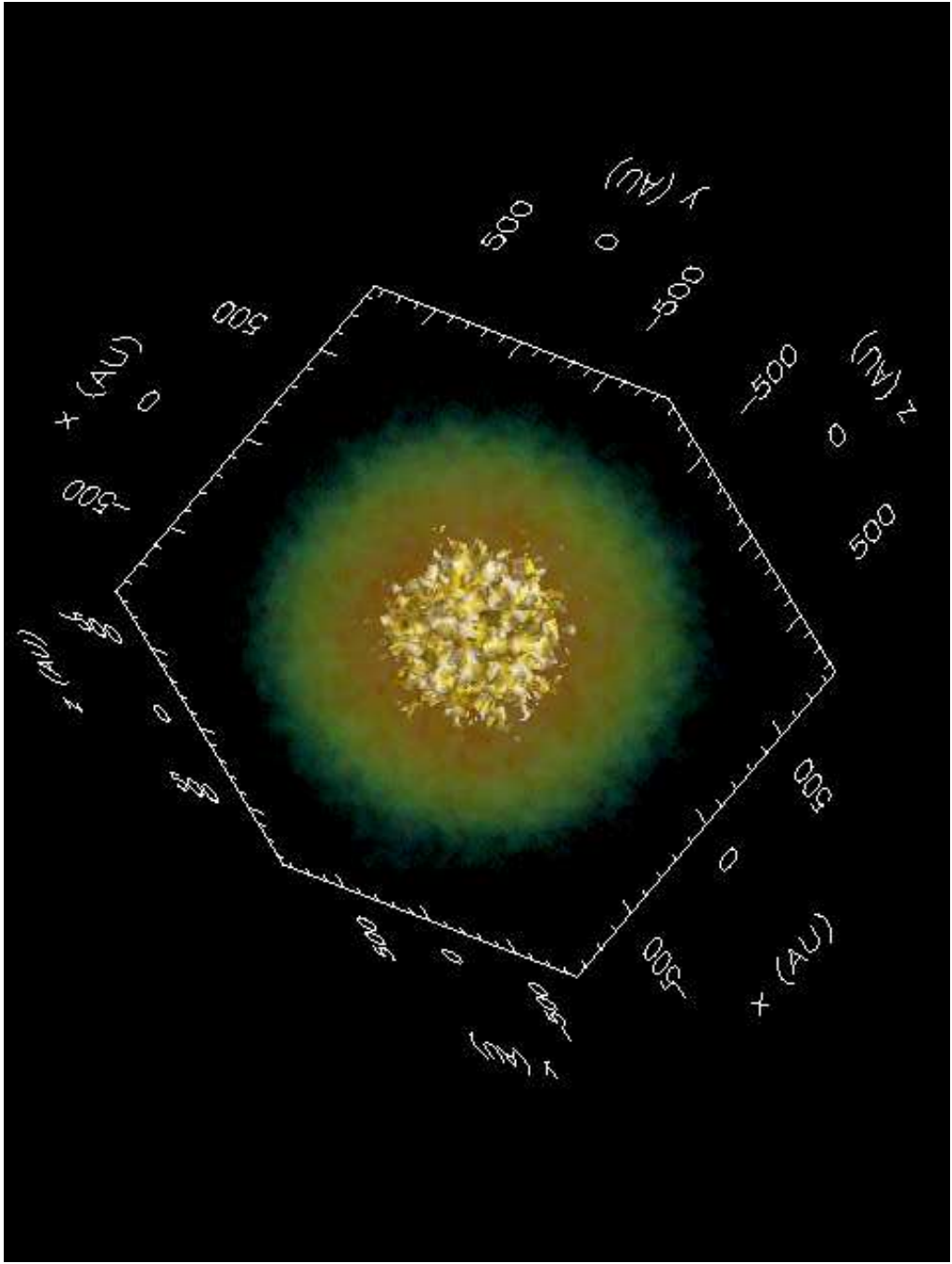}{7in}{0}{80}{80}{-250}{-10}
\label{fig:3dsym.ps}
\caption{3-dimensional simulation of the symmetric explosion 1 year
after the shock launch.  The contours represent the cobalt
distribution with a number density of $10^{-5}$.  The colors denote
the density distribution.  Note that although the explosion is
symmetric, Rayleigh-Taylor instabilities mix out the nickel.}
\end{figure}
\clearpage

\begin{figure} 
\plotfiddle{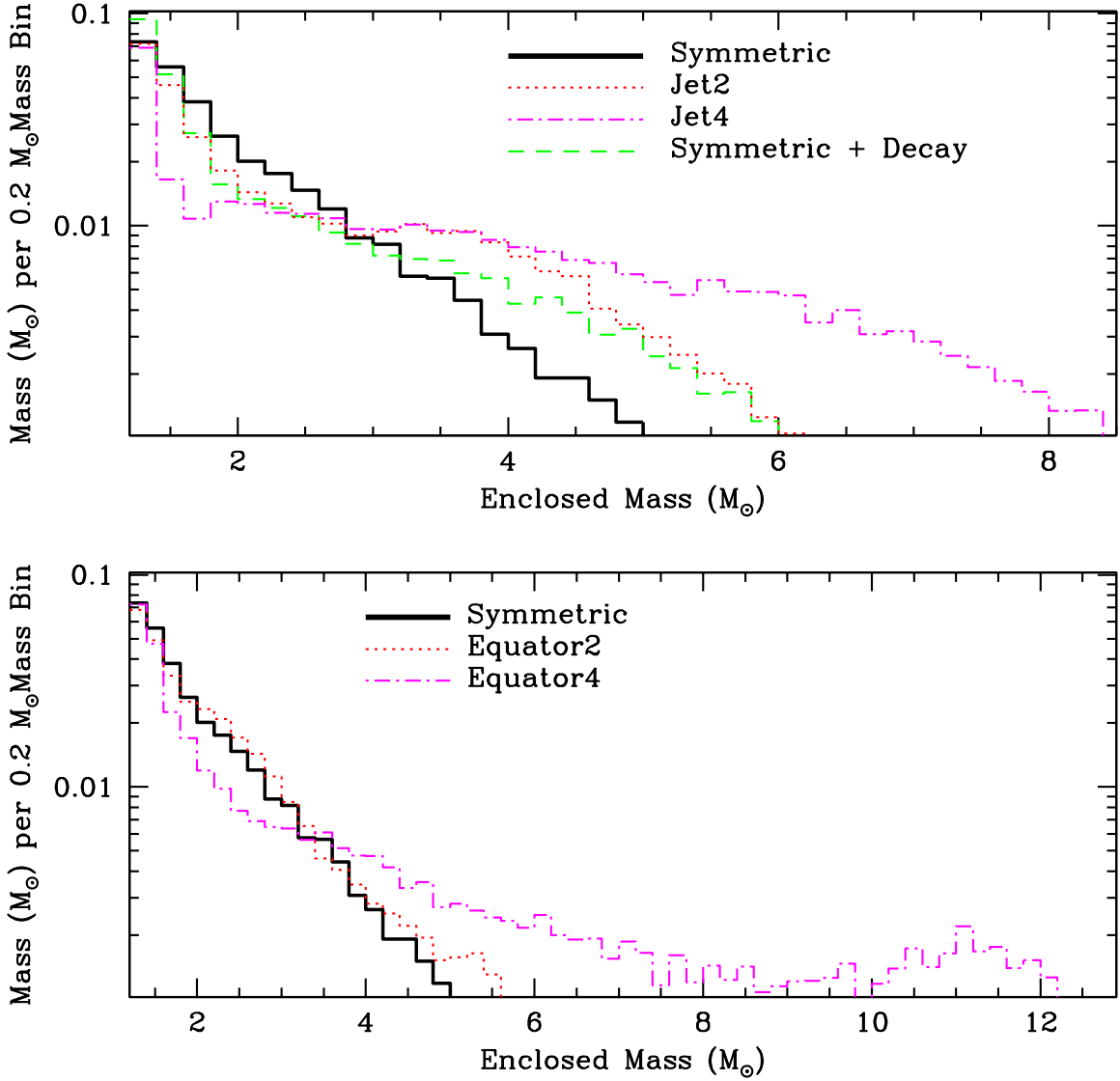}{7in}{0}{80}{80}{-250}{-50}
\label{fig:mvsm} 
\caption{Distribution of the nickel ejecta in mass, 
comparing jet explosions with a symmetric explosion (top panel) and
equatorial explosions with a symmetric explosion (bottom panel).  Note
that as we increase the degree of asymmetry (Jet4 versus Jet2,
Equator4 versus Equator2), the mixing increases dramatically, placing
nickel well into the hydrogen envelope of the star.  The dashed line
(top panel) shows the extent of mixing if all of the nickel/cobalt
decay energy is deposited into the nickel ejecta and produces almost 
as much mixing in a symmetric explosion as the Jet2 model.}
\end{figure}
\clearpage

\begin{figure}
\plotfiddle{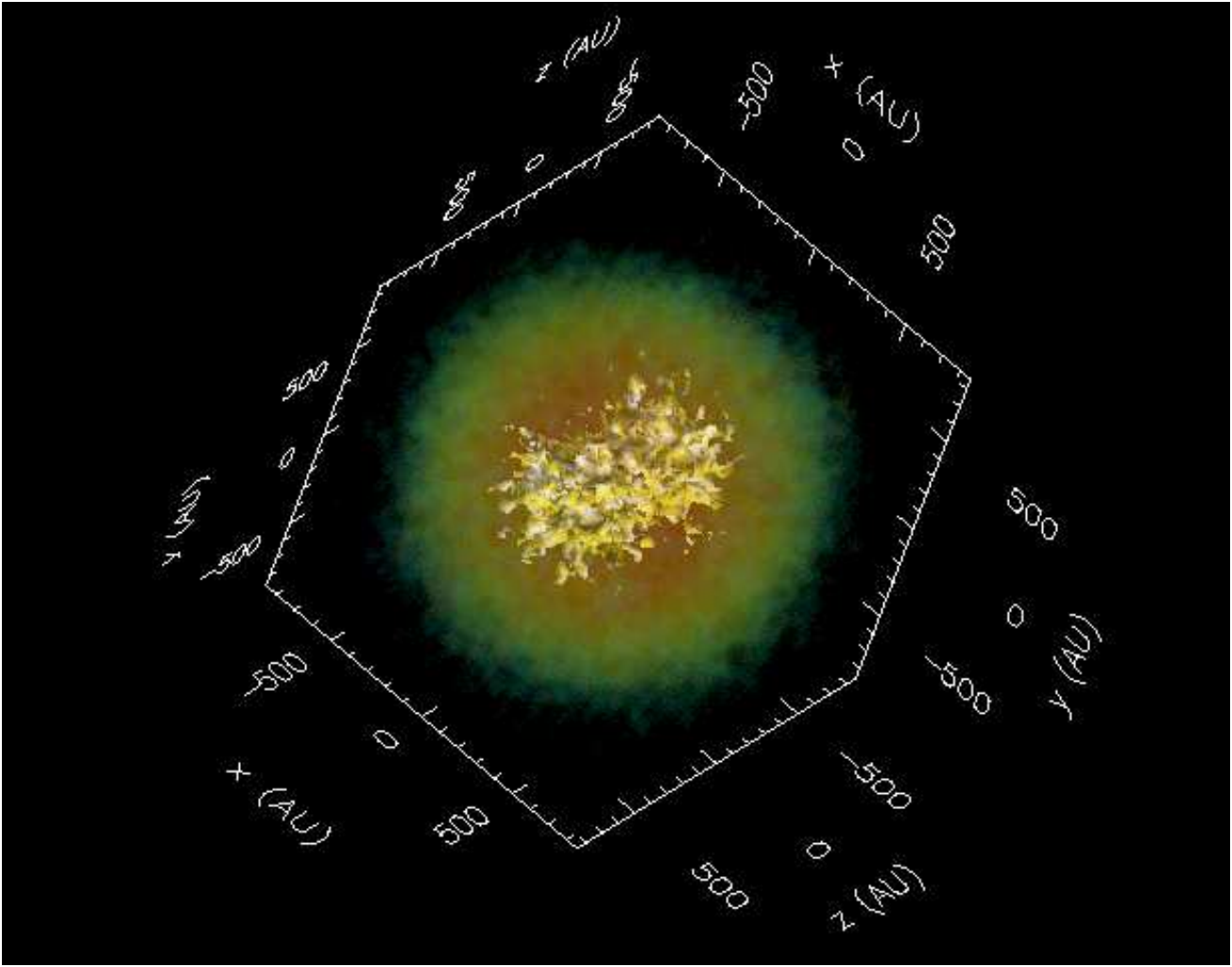}{7in}{0}{80}{80}{-250}{-10}
\label{fig:3dpole.ps}
\caption{3-dimensional simulation of the Jet2 asymmetric explosion 1
year after shock launch.  As in Fig. 3, the contours represent the
cobalt distribution with a number density of $10^{-5}$.  The colors
denote the density distribution.  The nickel is mixed out extensively
in the polar direction where the explosion was strongest.  However,
the density distribution did not gain large asymmetries and remained
fairly symmetric.}
\end{figure}
\clearpage

\begin{figure} 
\plotfiddle{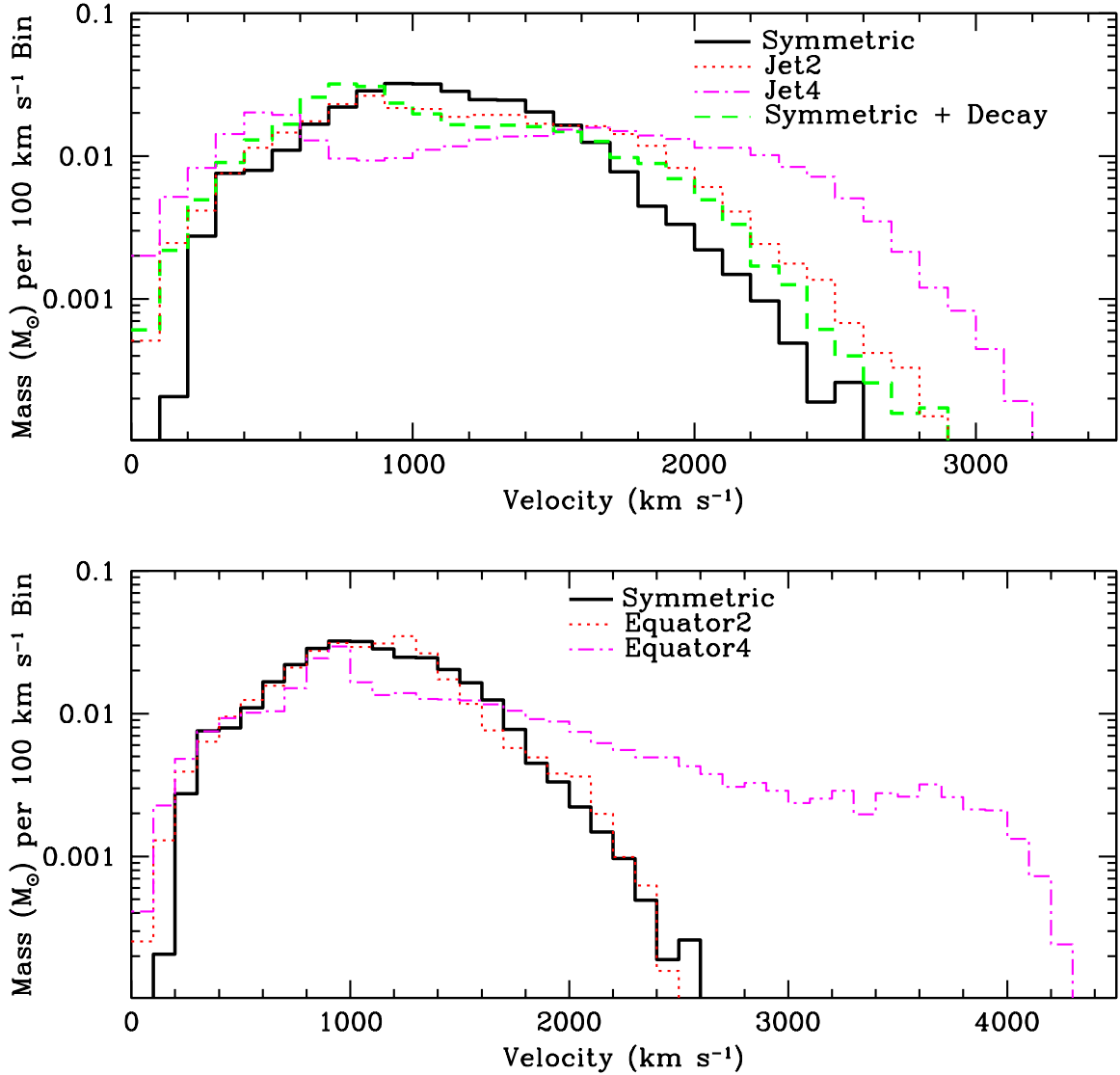}{7in}{0}{80}{80}{-250}{-50}
\label{fig:mvsv} 
\caption{Distribution of the nickel ejecta versus velocity
comparing jet explosions with a symmetric explosion (top panel) and
equatorial explosions with a symmetric explosion (bottom panel).  Note
that as we increase the degree of asymmetry (Jet4 versus Jet2,
Equator4 versus Equator2), the mixing increases dramatically, producing 
nickel velocities in excess of 3000\,km\,s$^{-1}$.  The dashed line
(top panel) shows the velocities achieved if all of the nickel/cobalt
decay energy is deposited into the nickel ejecta and produces almost 
as much mixing in a symmetric explosion as the Jet2 model.}
\end{figure}
\clearpage

\begin{figure} 
\plotfiddle{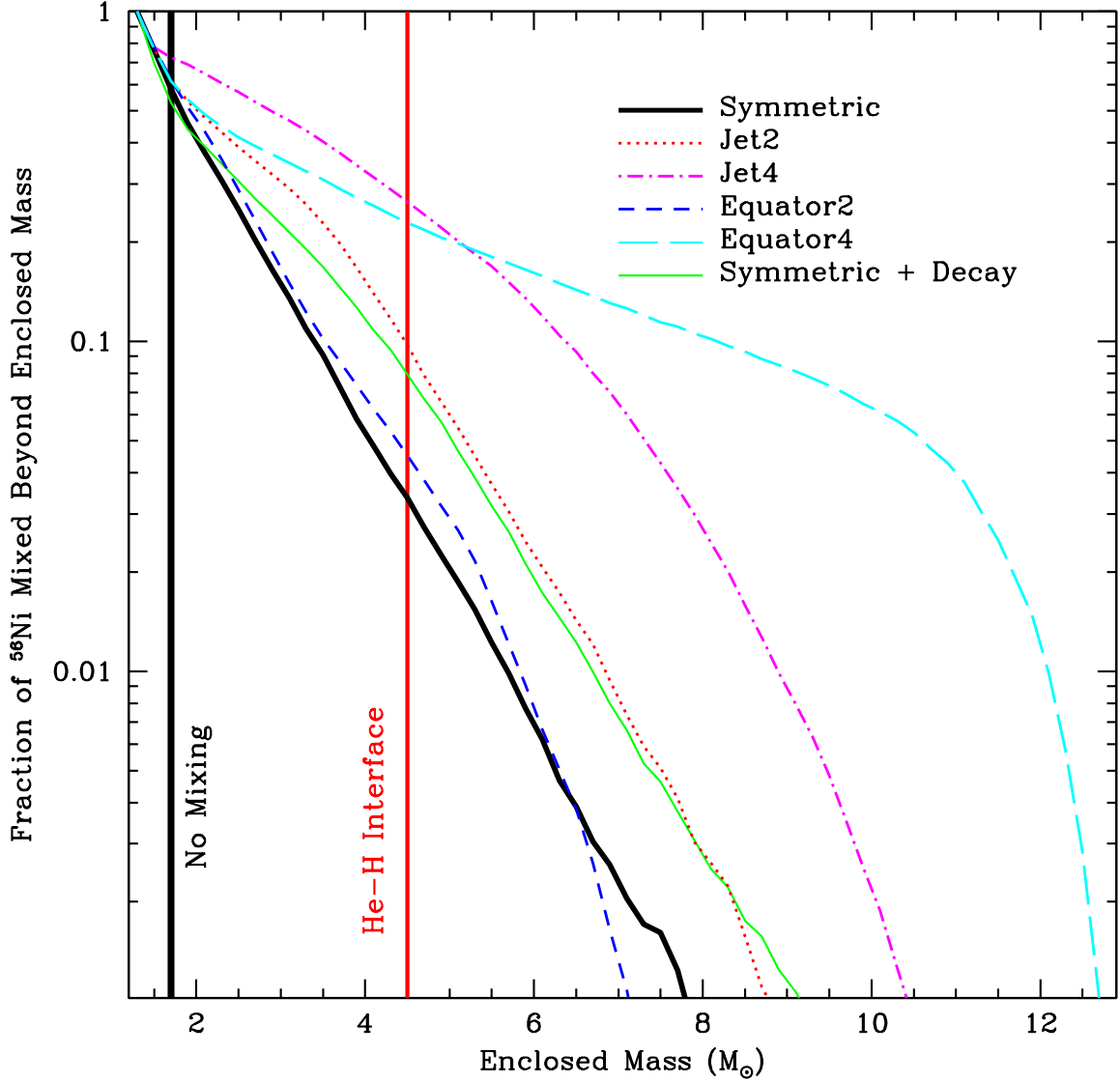}{7in}{0}{80}{80}{-250}{-50}
\label{fig:mixing} 
\caption{Fraction of nickel ejecta mixed out into the star for our set
of simulations.  Note that for mild asymmetries (Jet2) or if decay
energy is included in a symmetric explosion (Symmetric+Decay), nearly
10\% of the nickel mass is injected into the hydrogen envelope.  If
this amount of mixing occurs in weak explosions, an explosion that
resulted in a 4.5\,M\sun black hole remnant would still eject a moderate 
amount of nickel.}
\end{figure}
\clearpage

\begin{figure}
\plotfiddle{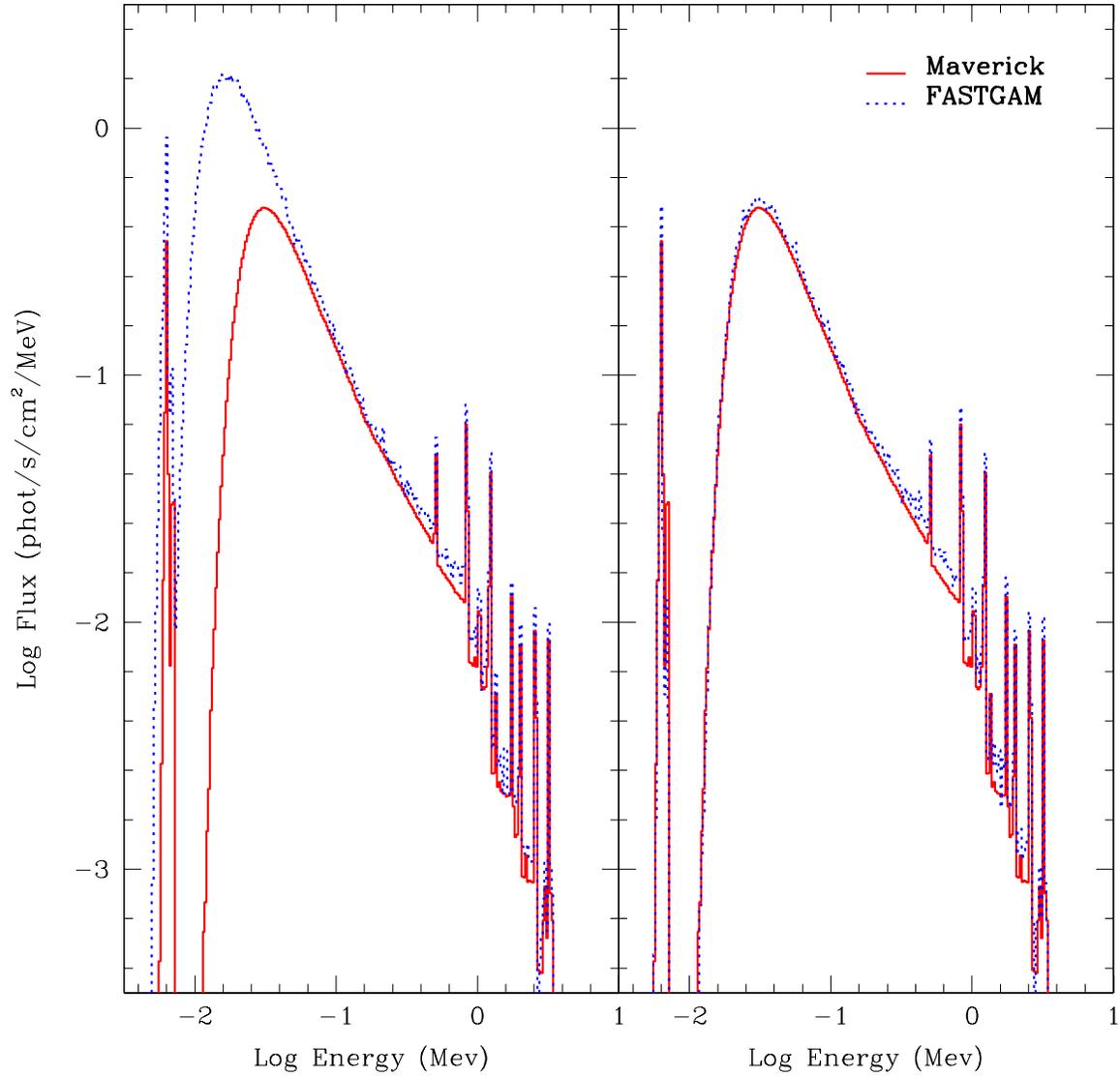}{7in}{0}{80}{80}{-250}{-50}
\label{fig:10hmmcompare.ps}
\caption{Total hard X- and $\gamma$-ray spectrum comparison
between the 1D Monte Carlo transport code FASTGAM and the 
3D code used in this work (Maverick).  The left panel shows
the two calculated spectra before a correction to the 
absorptive opacities in FASTGAM.  The right panel shows the
comparison once this correction had been made.  The agreement
between the two code results is quite good.}
\end{figure}
\clearpage

\begin{figure}
\plotfiddle{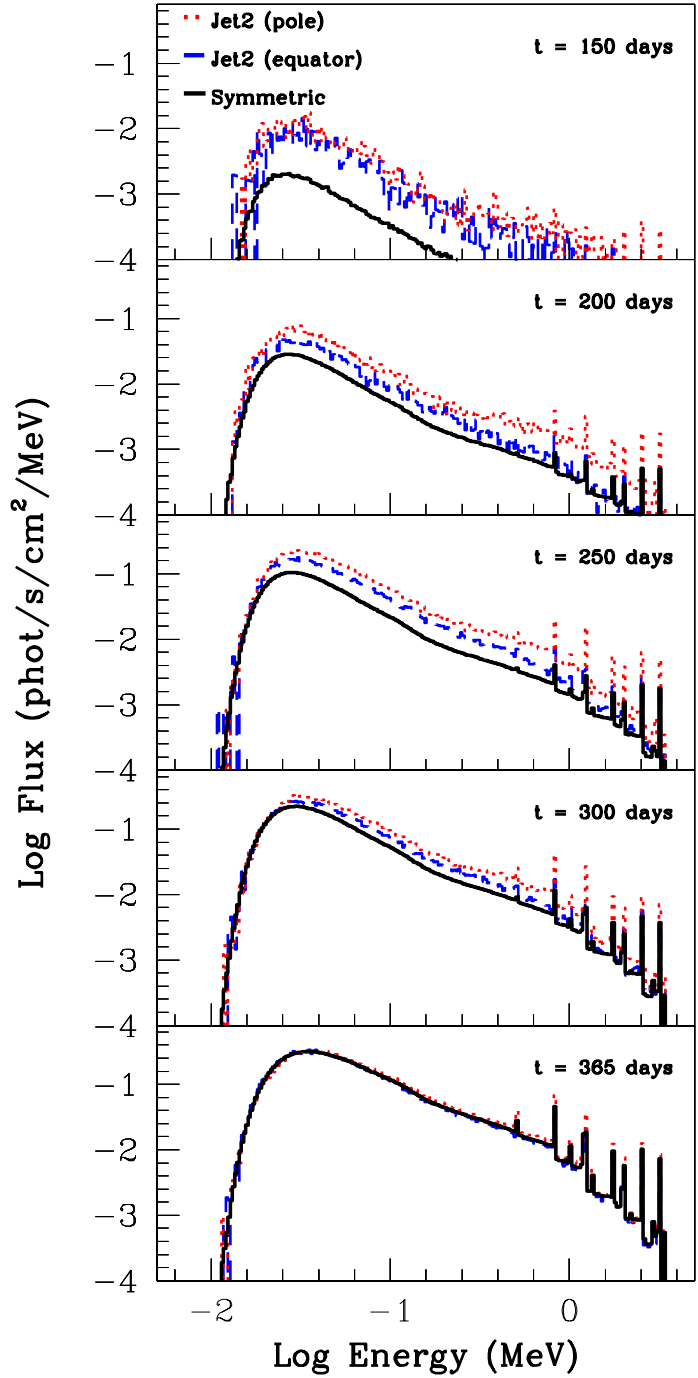}{7in}{0}{100}{100}{-170}{-160}
\label{fig:totspectime.ps}
\caption{Total hard X- and $\gamma$-ray spectrum at 5 different times
during the explosion (150,200,250,300,365 days) for  symmetric (solid
lines) and aspherical (Jet2) explosions (dotted and dashed lines).  The
flux is determined by assuming the object is 60\,kpc from the
observer.  The dotted lines refer to an aspherical explosion where the
jet is directed along the line-of-sight of the observer.  The dashed
lines refer to an explosion where the observer line-of-sight is directed 
$90^{\circ}$ off of the jet axis, in the equatorial direction.  Regardless
of observer viewing angle, the aspherical explosion is $\sim$2 brighter
than the symmetric explosion. }
\end{figure}
\clearpage

\begin{figure}
\plotfiddle{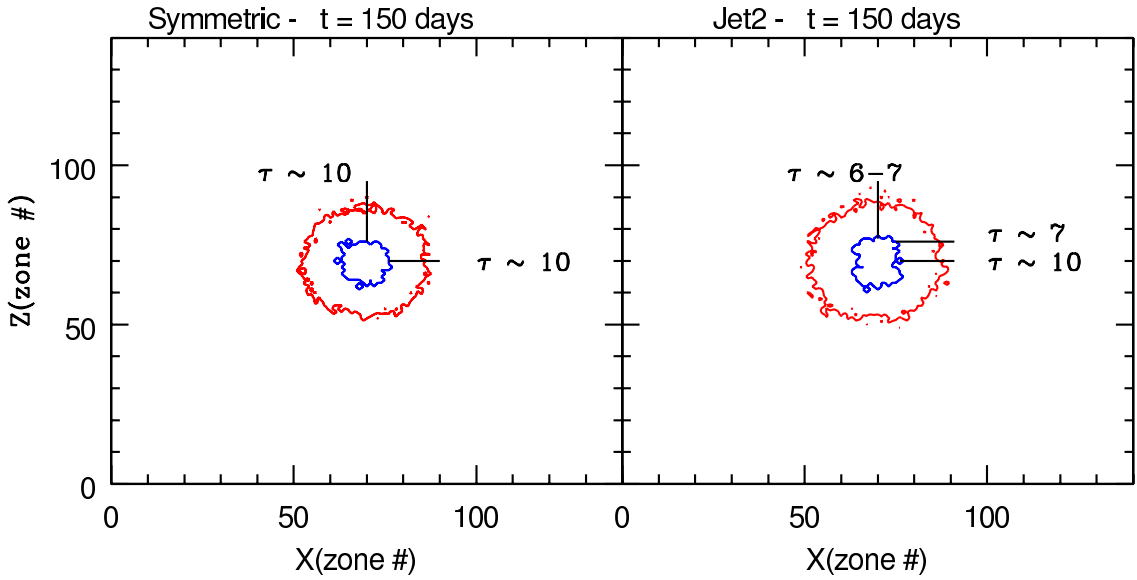}{7in}{0}{80}{80}{-250}{-10}
\label{fig:co_tau.ps_b}
\caption{Contour plots in the xz-plane of the Symmetric and Jet2
explosion models at t = 150 days.  Inner contour is for $^{56}$Co
number density which traces the surface of the $\gamma$-ray emitting
region.  Outer contour is for the mass density which follows electron
density and thus traces the dominant opacity source (Compton
scattering).  The lines represent lines-of-sight through the ejecta
for which the optical depth from emission region to ejecta surface has
been calculated.  Regardless of viewing angle, the optical depth of
the $^{56}$Co ejected along the poles in the Jet2 explosion remains
quite low.  Hence, it is this material that dominates the observed emission 
for all viewing angles in the aspherical explosion.}
\end{figure}
\clearpage

\begin{figure}
\plotfiddle{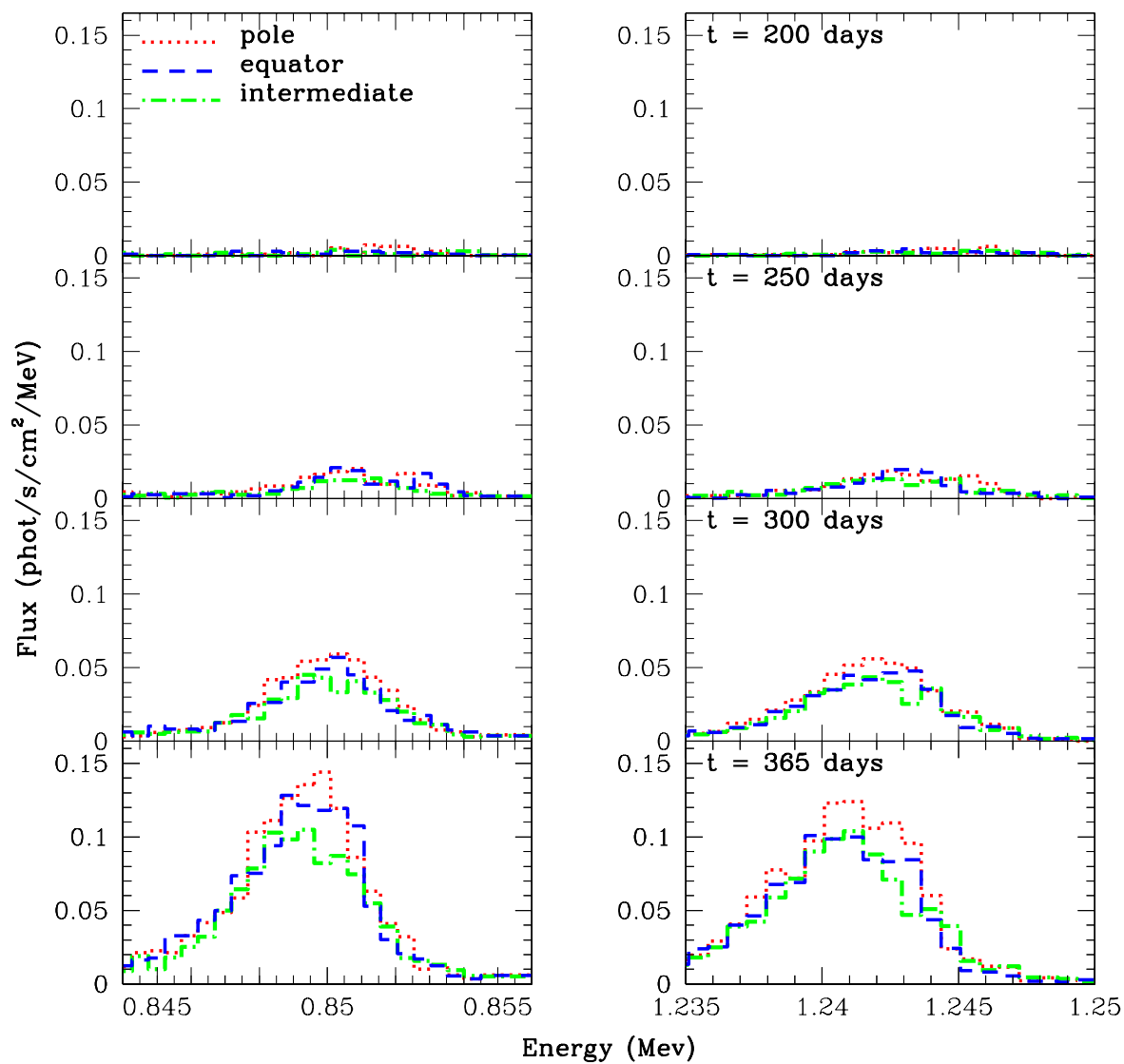}{7in}{0}{80}{80}{-250}{-50}
\label{fig:lineproftimesymm.ps}
\caption{Line profiles of the $^{56}$Co 1.238 and 0.847 MeV lines for
the Symmetric model at 4 different times during the explosion (200,
250, 300, 365 days).  3 different viewing angles are shown: polar view
(dotted lines), equatorial view (dashed lines) and an intermediate
view angle of $\sim~45^{\circ}$ (dash-dot lines).  The line profiles
do not show significant variation with viewing angle (as would be
expected for a symmetric explosion.)  }
\end{figure}
\clearpage

\begin{figure}
\plotfiddle{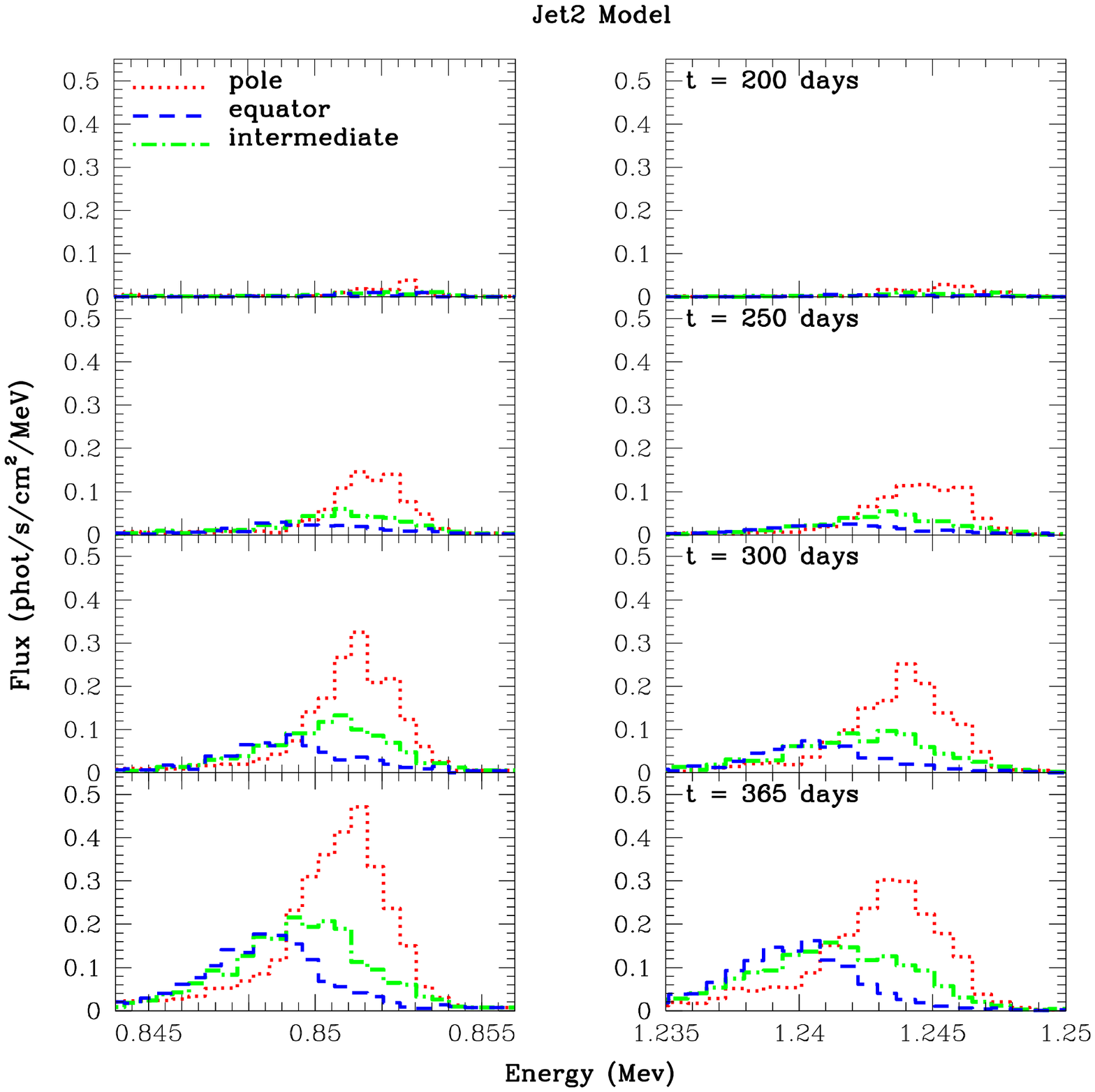}{7in}{0}{80}{80}{-250}{-50}
\label{fig:lineproftime.ps}
\caption{Line profiles of the $^{56}$Co 1.238 and 0.847 MeV lines for
the Jet2 model at 4 different times during the explosion (200, 250,
300, 365 days).  3 different viewing angles are shown: polar view
(dotted lines), equatorial view (dashed lines) and an intermediate
view angle of $\sim~45^{\circ}$ (dash-dot lines).  The flux axis is 
scaled by a factor of 4 over the Symmetric model profiles shown
in Figure 11.  Significant variations in the line profiles with viewing
angle are apparent, and can be explained by considering the velocity
distribution (and thus radial distribution in a homologous expansion) of 
the ejecta responsible for the observed emission.} 
\end{figure}

\begin{figure}
\plotfiddle{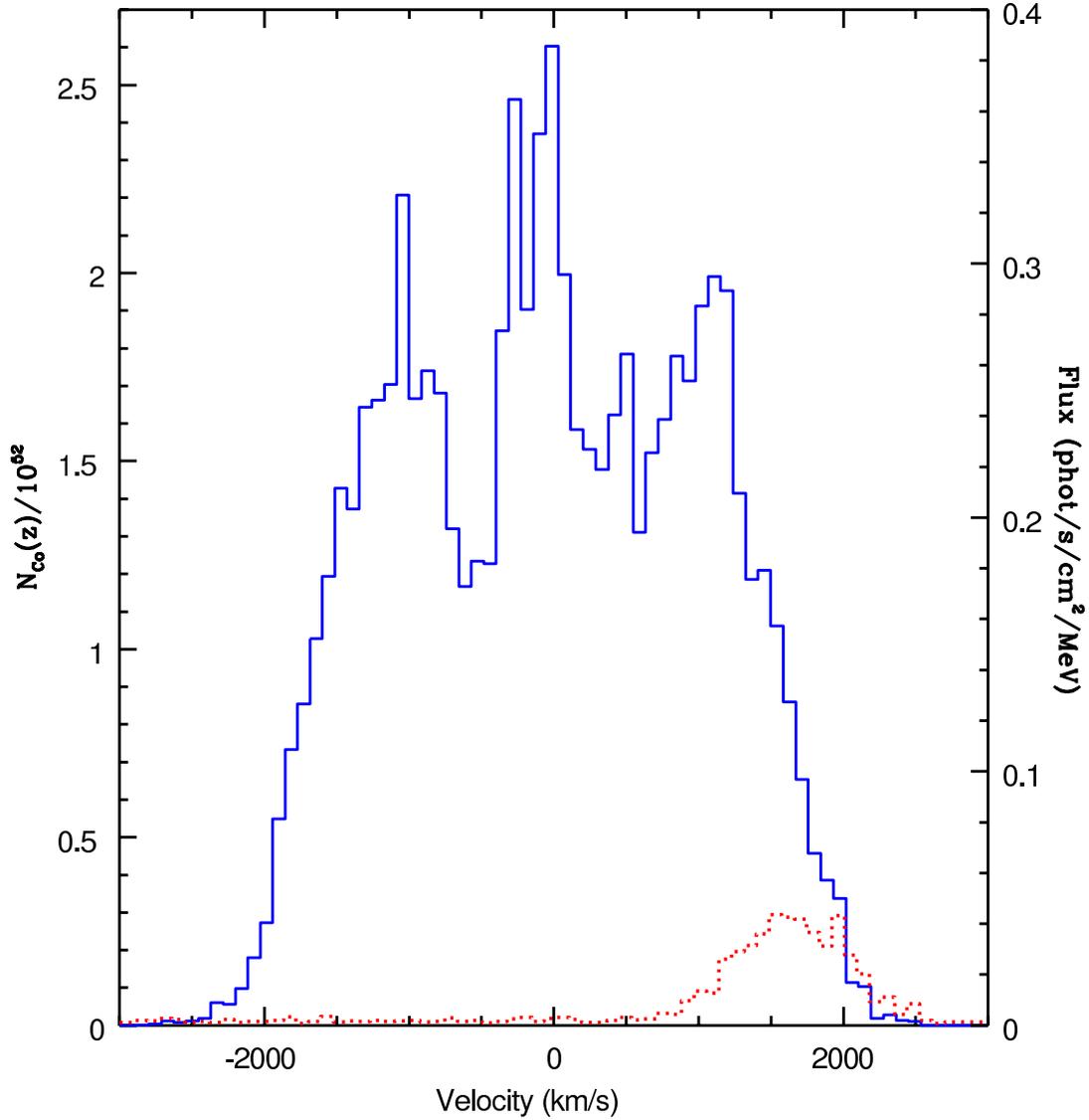}{7in}{0}{80}{80}{-250}{-50}
\label{fig:nidistprof.ps}
\caption{Number of cobalt nuclei (${\mathbf N}_{\mathrm Co}(z)$)
versus ejecta velocity along the polar axis for the Jet2 model at 250
days with roughly 3 times more spatial resolution than the previous
run.  The $\gamma$-ray line flux for this polar view is plotted as the
dotted line for comparison.  Positive velocities correspond to
blueshifted energies.  The line flux at this epoch departs from the
cobalt distribution at velocities of order 1800~km/s due to optical
depth effects.  As the supernova expands the line profile will take on
the irregular shape of the underlying distribution.}
\end{figure}
\clearpage

\end{document}